\documentclass[english,superscriptaddress,amssymb,twocolumn]{revtex4}
\newcommand{\bh}[1]{#1}

\usepackage{bm}
\usepackage{graphicx}
\usepackage{color}
\usepackage{amsmath}

\usepackage[T1]{fontenc}
\usepackage[latin9]{inputenc}
\usepackage{textcomp}
\usepackage{amssymb}
\usepackage{ulem}
\usepackage{extraplaceins}
\makeatother

\usepackage[english]{babel}

\begin{document}

\title{Observing quantum state diffusion by heterodyne detection of fluorescence}

\author{P. Campagne-Ibarcq}

\affiliation{Laboratoire Pierre Aigrain, Ecole Normale Sup\'erieure-PSL Research University, CNRS, Universit\'e Pierre et Marie Curie-Sorbonne Universit\'es, Universit\'e Paris Diderot-Sorbonne Paris Cit\'e, 24 rue Lhomond, 75231 Paris Cedex 05, France}
\affiliation{Quantic Team, INRIA Paris-Rocquencourt, Domaine de Voluceau, B.P. 105, 78153 Le Chesnay Cedex, France}

\author{P. Six}

\affiliation{Centre Automatique et Syst\`emes, Mines ParisTech, PSL Research University,
60 Boulevard Saint-Michel, 75272 Paris Cedex 6, France.}
\affiliation{Quantic Team, INRIA Paris-Rocquencourt, Domaine de Voluceau, B.P. 105, 78153 Le Chesnay Cedex, France}

\author{L. Bretheau}

\affiliation{Laboratoire Pierre Aigrain, Ecole Normale Sup\'erieure-PSL Research University, CNRS, Universit\'e Pierre et Marie Curie-Sorbonne Universit\'es, Universit\'e Paris Diderot-Sorbonne Paris Cit\'e, 24 rue Lhomond, 75231 Paris Cedex 05, France}
\affiliation{Quantic Team, INRIA Paris-Rocquencourt, Domaine de Voluceau, B.P. 105, 78153 Le Chesnay Cedex, France}

\author{A. Sarlette}

\affiliation{Quantic Team, INRIA Paris-Rocquencourt, Domaine de Voluceau, B.P. 105, 78153 Le Chesnay Cedex, France}

\author{M. Mirrahimi}

\affiliation{Quantic Team, INRIA Paris-Rocquencourt, Domaine de Voluceau, B.P. 105, 78153 Le Chesnay Cedex, France}

\author{P. Rouchon}

\affiliation{Centre Automatique et Syst\`emes, Mines ParisTech, PSL Research University,
60 Boulevard Saint-Michel, 75272 Paris Cedex 6, France.}
\affiliation{Quantic Team, INRIA Paris-Rocquencourt, Domaine de Voluceau, B.P. 105, 78153 Le Chesnay Cedex, France}

\author{B. Huard}
\email{benjamin.huard@ens.fr}
\affiliation{Laboratoire Pierre Aigrain, Ecole Normale Sup\'erieure-PSL Research University, CNRS, Universit\'e Pierre et Marie Curie-Sorbonne Universit\'es, Universit\'e Paris Diderot-Sorbonne Paris Cit\'e, 24 rue Lhomond, 75231 Paris Cedex 05, France}
\affiliation{Quantic Team, INRIA Paris-Rocquencourt, Domaine de Voluceau, B.P. 105, 78153 Le Chesnay Cedex, France}

\date{\today}
\begin{abstract}
A qubit can relax by fluorescence, which prompts the release of a photon into its electromagnetic environment. By counting the emitted photons, discrete quantum jumps of the qubit state can be observed. The succession of states occupied by the qubit in a single experiment, its quantum trajectory, depends in fact on the kind of detector. How are the quantum trajectories modified if one measures continuously the amplitude of the fluorescence field instead? Using a superconducting parametric amplifier, we have performed heterodyne detection of the fluorescence of a superconducting qubit. For each realization of the measurement record, we can reconstruct a different quantum trajectory for the qubit. The observed evolution obeys quantum state diffusion, which is characteristic of quantum measurements subject to zero point fluctuations. Independent projective measurements of the qubit at various times provide a quantitative validation of the reconstructed trajectories. By exploring the statistics of quantum trajectories, we demonstrate that the qubit states span a deterministic surface in the Bloch sphere at each time in the evolution. Additionally, we show that when monitoring fluorescence, coherent superpositions are generated during the decay from excited to ground state. Counterintuitively, measuring light emitted during relaxation can give rise to trajectories with increased excitation probability.
\end{abstract}

\maketitle

\section{Introduction}

The quantum properties of an open system are preserved as long as no information about its state is lost into unmonitored degrees of freedom~\cite{zurek2015classical}. For a qubit, the mere possibility to emit a photon by spontaneous emission can lead to decoherence. When discarding the information carried by the emitted field, the imperfectly known qubit state is described by a density matrix, which evolves continuously towards the ground state. If instead an observer monitors the fluorescence light emitted by the qubit, the lost information is retrieved. The succession of states occupied by the qubit then depends on the particular realization of the random measured record and deviates from the average one~\cite{carmichael1993quantum,wiseman2009quantum,barchielli2009}. 

Peculiar to quantum mechanics, this quantum trajectory depends on the type of detection. In case of photocounting, the qubit would undergo discrete quantum jumps. In contrast, Wiseman and Milburn showed in 1992 that heterodyne measurement of fluorescence should lead to continuous quantum state diffusion~\cite{Gisin1992,Wiseman1993}. 
Experimentally, quantum jumps have been observed in a variety of physical systems~\cite{carmichael1993quantum,wiseman2009quantum,barchielli2009,Carmichael2009}, not by directly measuring the emitted photon during the system decay, but by using the information extracted from an ancilla, such as extra energy levels~\cite{Bergquist1986,Nagourney1986,Sauter1986,Basche1995}, coupled qubits~\cite{Gleyzes2007,Yang2008,Vamivakas2010,Neumann2010,Vool2014,Sun2014} or harmonic oscillators~\cite{Peil1999,Vijay2011}. In turn, diffusive quantum trajectories were recently observed in a superconducting qubit by extracting information from an ancilla, through the continuous measurement of the quadratures of a coupled microwave mode~\cite{murch2013observing,hatridge2013quantum,murch2014,DeLange2014,Weber2015}.
Here, we perform a direct heterodyne measurement of the light emitted during qubit decay without any ancillary system in the original spirit of Ref.~\cite{Wiseman1993}. We perform such a measurement on a superconducting qubit using a phase-preserving parametric amplifier \cite{bergeal2010phase, Roch2012}, which ensures an overall measurement efficiency $\eta=24~\%$ for the qubit relaxation channel. Without drive and starting from an initially pure state, we demonstrate that the qubit state evolves erratically towards the ground state $|g\rangle$, in agreement with quantum state diffusion~\cite{Wiseman1993}. The validity of the obtained quantum trajectories is verified using an independent tomographic measurement based on an ancillary oscillator. Counterintuitively, the sole monitoring of relaxation can temporarily increase the probability amplitude of excitation~\cite{Bolund2014}. It can also generate coherences out of energy eigenstates, which is in contrast with the recently observed quantum trajectories based on quantum non demolition measurement of the qubit~\cite{murch2013observing,hatridge2013quantum,murch2014,DeLange2014,Weber2015}.

\begin{figure}
\includegraphics[scale=.45]{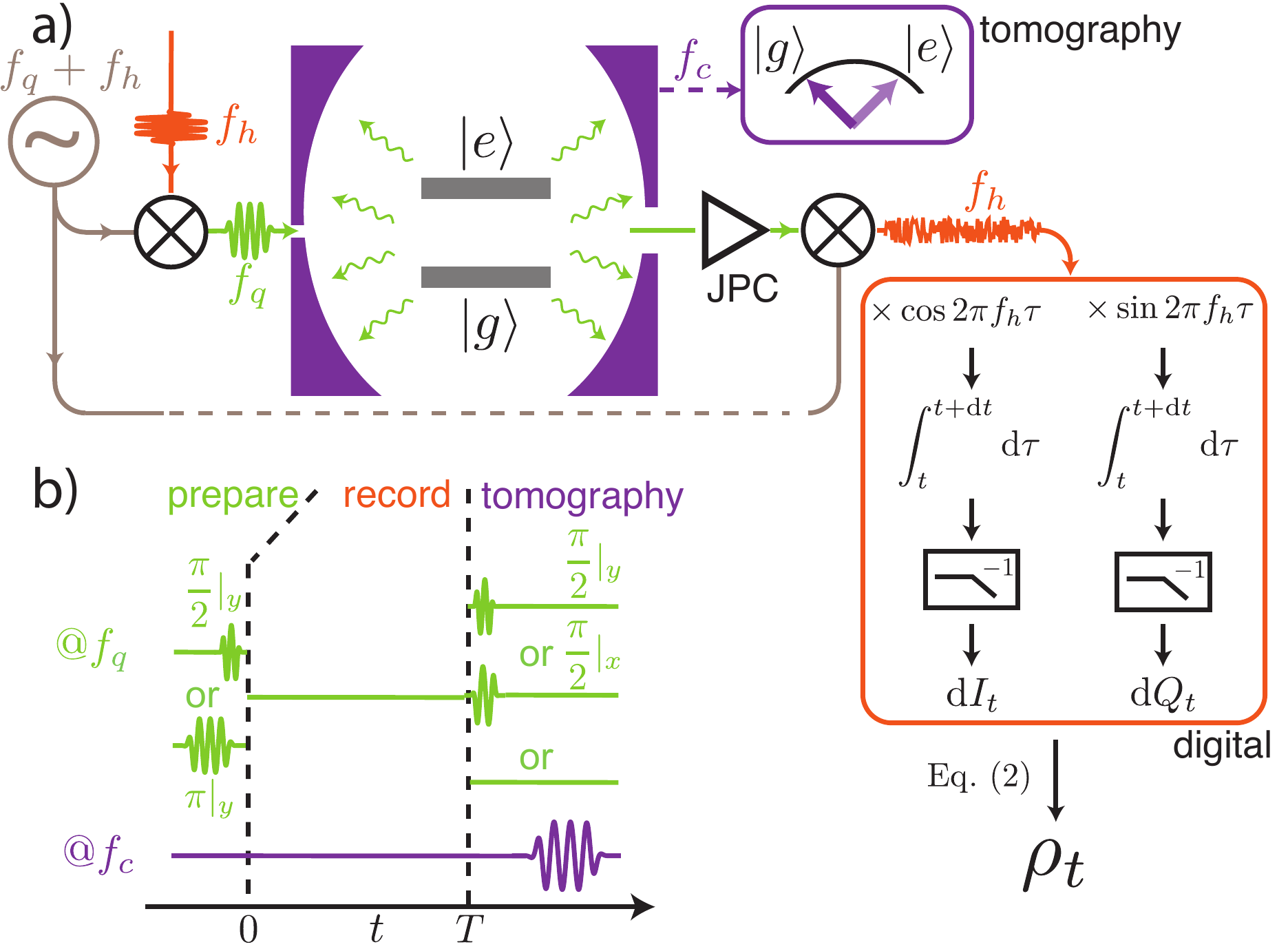}
\caption{\label{fig1} \textbf{Scheme of the experiment. a)} The fluorescence field of a superconducting qubit in an off-resonant cavity is recorded using a heterodyne detection setup from time $0$ to $T$. Following amplification by a Josephson Parametric Converter (JPC), the signal is downconverted to $f_h=100~\mathrm{MHz}$ and numerically demodulated and integrated on time steps $\mathrm{d}t= 200~\mathrm{ns}$ into its two quadratures $\mathrm{d}I$ and $\mathrm{d}Q$. The rectangular symbols represent the correction of finite detection bandwidth in the setup~\cite{supmat}. The quantum trajectory $\{\rho_t\}$ is then computed using Eq.~\eqref{SME}. A single local oscillator at $f_q+f_h$ is used for qubit manipulation and downconversion of the fluorescence signal. The transmission at cavity frequency $f_c=7.8~\mathrm{GHz}$ is used to independently readout the qubit at time $T$~\cite{supmat}. \textbf{b)} Pulse sequence. At time $0$, the qubit is prepared in $|+x\rangle$ (resp. $|+e\rangle$) with a 52~ns (resp. 104~ns) rotation pulse around $\sigma_y$. The fluorescence record is acquired for a duration $T$ ranging from 0 to $10~\mu\mathrm{s}$. The state is then projectively readout along one of the three Pauli operators using a pulse at frequency $f_c$ preceded by a rotation pulse around $\sigma_y$, $\sigma_x$, or no pulse.}
\end{figure}

In the experiment sketched in Fig.~\ref{fig1}a, the qubit under monitoring is a transmon resonating at $f_q=6.37~\mathrm{GHz}$. It is dispersively coupled to a 3D bulk copper cavity~\cite{Koch2007,Paik2011}, which serves two purposes. First, it channels most of the emission coming from qubit relaxation into a dominantly coupled output transmission line~\cite{Purcell46}. The qubit decay rate is measured to be $\gamma_1=(4.15~\mu\mathrm{s})^{-1}$. Second, it can be used to perform a projective readout of the qubit~\cite{blais2004cavity,supmat} that will be used as a validation of the quantum trajectories in section~\ref{sectionvalidation}. At time $t=0$, the qubit is prepared either in the excited state $|e\rangle$ or in $|+x\rangle=\frac{1}{\sqrt{2}}(|g\rangle+|e\rangle)$. This is done by applying a rotation pulse (Fig.~\ref{fig1}b) on the qubit initially at equilibrium, where its excitation is below $1~\%$ consistently with dilution refrigerator temperatures. The qubit is then left to decay while heterodyne detection of the fluorescence field is performed on the output line using a high efficiency detection setup based on a Josephson Parametric Converter (JPC)~\cite{Bergeal2010,Roch2012,supmat}.

  \begin{figure}
\includegraphics[scale=.25]{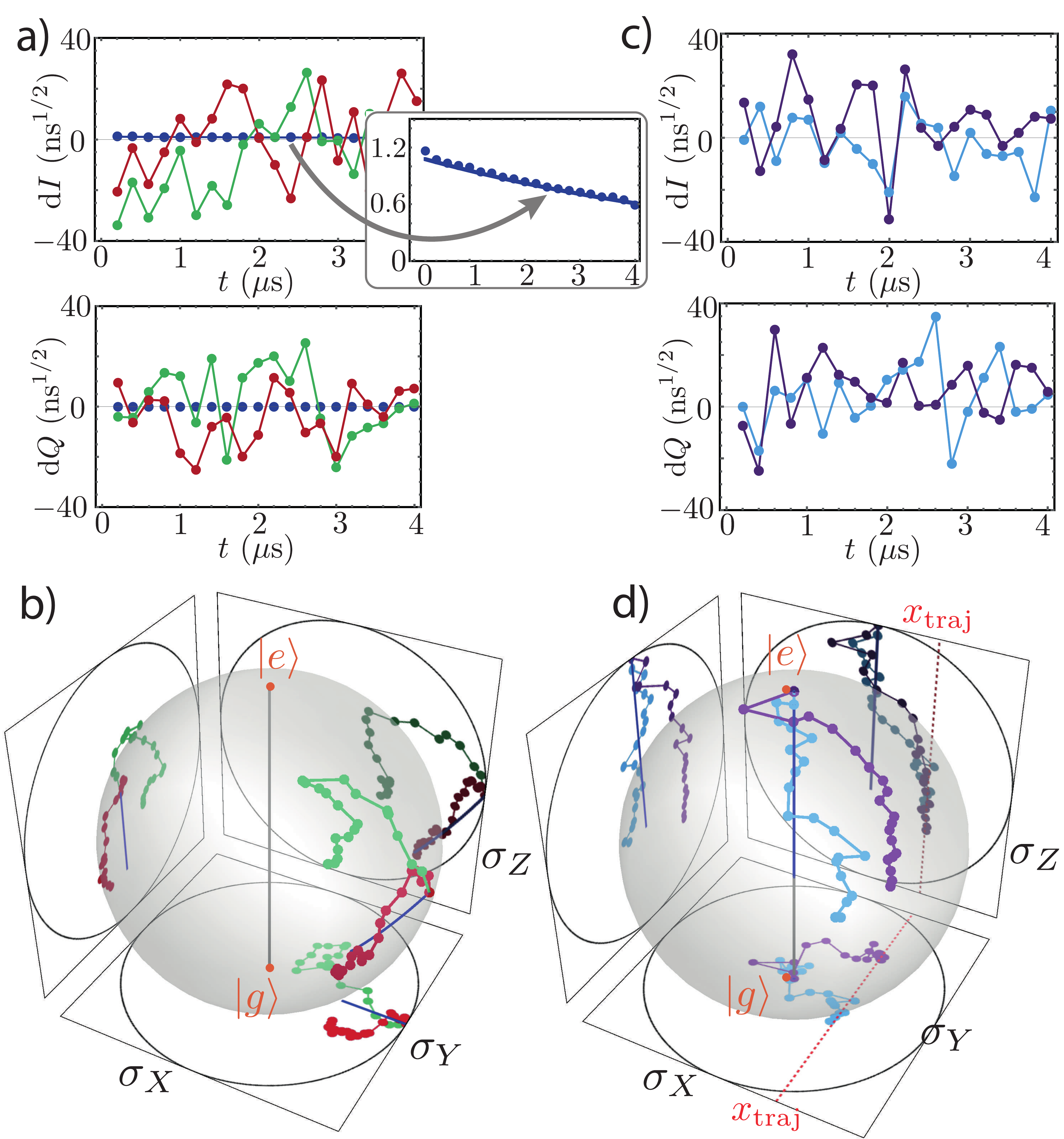}
\caption{\label{fig2} \textbf{Quantum trajectories.} \textbf{a)}  Each panel displays two different measurement records of one quadrature of the fluorescence field for a qubit initially in $|+x\rangle$. Actual measurements are shown as red and green dots linked by straight lines for clarity. The blue dots correspond to the average record on all experiments, and the blue line corresponds to an exponential fit of these dots in $e^{-\gamma_1t/2-\gamma_\phi t}$. A zoom in is shown as an inset. Measurements of $\mathrm{d}I$ and $\mathrm{d}Q$ are normalized by an overall prefactor so that their variance is $\mathrm{d}t$~\cite{supmat}. \textbf{b)} Quantum trajectories followed by the qubit under relaxation represented in the Bloch sphere for the measurement records shown in Fig.~\ref{fig2}a. The Bloch vector $(x,y,z)$ corresponds to the state $\rho=(\mathbf{1}+x\sigma_x+y\sigma_y+z\sigma_z)/2$ and the black circles set the scale of the Bloch sphere extrema. The trajectories are obtained by solving numerically Eq.~\eqref{SME} and are shown as dots linked by straight lines both in the sphere and in the three projections along $x$, $y$ and $z$. Colors identify which record of Fig.~\ref{fig2}a is used. A blue line shows the average evolution without monitoring ($\eta=0$). \textbf{c,d)} Same representations for two realizations starting from $|e\rangle$. They were arbitrarily selected to end up in a state with similar Bloch coordinate $x_{\mathrm{traj}}=0.42 \pm 0.02$ as indicated by a dashed red line.}
\end{figure}

\section{Computing quantum trajectories from heterodyne measurement of fluorescence}

Previous experiments on superconducting circuits have shown that the fluorescence field contains a footprint of the qubit state~\cite{Houck2007,astafiev2010resonance,Abdumalikov2011,campagne2014observing}.  Indeed, the integrated outputs $\mathrm{d}I_t$ and $\mathrm{d}Q_t$ of an heterodyne detector between times $t$ and $t+\mathrm{d}t$ are on average proportional to $\langle\sigma_x\rangle_{\rho_t}=\mathrm{Tr}(\rho_t\sigma_x)$ and $\langle\sigma_y\rangle_{\rho_t}$ respectively, where $\rho_t$ is the density matrix of the qubit and $\sigma_{x,y,z}$ the Pauli operators. In the inset of Fig.~\ref{fig2}a is shown in blue the average of 3 millions measurement records $\{\mathrm{d}I_t,\mathrm{d}Q_t\}_{0<t<T}$ for a qubit starting in $|+x\rangle$. As expected, the quadrature $\mathrm{d}I_t$ decays exponentially at a rate $\gamma_1/2+\gamma_\phi$ while the quadrature $\mathrm{d}Q_t$ is zero. We note that the pure dephasing rate $\gamma_{\phi}\approx(35~\mu\mathrm{s})^{-1}$ is measured to be much smaller than $\gamma_1$. Two individual measurement records are shown in Fig.~\ref{fig2}a for a qubit starting in $|+x\rangle$ and in Fig.~\ref{fig2}c when starting in $|e\rangle$. They are fluctuating with a much larger amplitude than the average signal. This noise originates both from zero point fluctuations of the detected field quadratures, which contain information on the qubit state, and from imperfections of the detection setup. The relative contribution of the first one is characterized by the efficiency $\eta$. Using the record $\{\mathrm{d}I_t,\mathrm{d}Q_t\}_{0<t<T}$ and the initial state $\rho_0$, one can reconstruct a quantum trajectory $\{\rho_t\}$ over the time $T$. The density matrix $\rho_t$ is here conditioned to the knowledge of the initial state at time $0$ and of the fluorescence record between $0$ and $t$.

Mathematically, the measurement records can be decomposed as~\cite{wiseman2009quantum} 
 \begin{equation}
 \left\{
 \begin{array}{lcl}
 \mathrm{d}I_t&=&\sqrt{\frac{\eta \gamma_1}{2}} \langle\sigma_x\rangle_{\rho_t} \mathrm{d}t+\mathrm{d}W_{I,t}\\
  \mathrm{d}Q_t&=&\sqrt{\frac{\eta \gamma_1}{2}} \langle\sigma_y\rangle_{\rho_t} \mathrm{d}t+\mathrm{d}W_{Q,t}
 \end{array}
 \right. .\label{dIdQ}
 \end{equation}
where $\mathrm{d}W_{I,t}$ and $\mathrm{d}W_{Q,t}$ are the random fluctuations beyond the expected average value that were discussed above. The signals are normalized so that the variance of $\mathrm{d}I_t$ and $\mathrm{d}Q_t$ is directly $\mathrm{d}t$ (see Ref.~\cite{supmat}  for details). The overall efficiency $\eta=24~\%$ is determined using a max-like method~\cite{six2015parameter}. It is limited by both the extra relaxation mechanisms that do not lead to emission into the output line and limited efficiency of the detection setup. 
The evolution of the density matrix can be inferred from these measurement records using the following Stochastic Master Equation. In the frame rotating at $f_q$, and without qubit drive, it reads~\cite{Gambetta2007,wiseman2009quantum}
  \begin{equation}
  \begin{array}{lccrcl}
  \mathrm{d}\rho_t&=&& \left(\gamma_1~ \mathcal{D}[\sigma_-]\rho_t+\gamma_{\phi}/2~\mathcal{D}[ \sigma_z]\rho_t\right)&\times&\mathrm{d} t\\
  &&&+ ~\sqrt{ \eta\gamma_1/2}~ \mathcal{M}[\sigma_-]\rho_t&\times&\mathrm{d}W_{I,t}\\
  &&&+\sqrt{ \eta\gamma_1/2} \mathcal{M}[i\sigma_-]\rho_t&\times&\mathrm{d}W_{Q,t}  ,\end{array}
  \label{SME}
    \end{equation} 
using the Lindblad $\mathcal{D}[L]\rho=L\rho L^{\dagger}-\frac{1}{2} L^{\dagger}L\rho-\frac{1}{2}\rho L^{\dagger}L$ and measurement $\mathcal{M}[L]\rho=(L-\langle L\rangle_{\rho})\rho + \rho(L-\langle L\rangle_{\rho})^{\dagger}$ superoperators and the lowering operator $\sigma_-=|g\rangle\langle e|$.  

When fluorescence is not monitored, which can be modeled by setting the measurement efficiency to zero, the qubit state dynamics is captured by the deterministic Lindblad terms. The first one corresponds to the average effect of relaxation, while the second one models pure dephasing and is almost negligible here. The corresponding evolution of the Bloch vector is plotted in blue in Figs.~\ref{fig2}b,d. 

By monitoring fluorescence during relaxation ($\eta>0$), the observer retrieves part of the information lost in the environment. The acquired information is injected in Eq.~\eqref{SME} solely via the noise terms $\mathrm{d}W_{I,t}$ and $\mathrm{d}W_{Q,t}$. The associated stochastic quantum backaction is captured by $\mathcal{M}[\sigma_-]$ and $\mathcal{M}[i \sigma_-]$ of Eq.~(\ref{SME}) (see \cite{Hofmann1998,supmat} for a graphical representation of backaction in the Bloch sphere). It is then possible to reconstruct the quantum trajectory of the qubit corresponding to a measurement record. In practice, we choose a time step $\mathrm{d}t=200~\mathrm{ns}$ close to the autocorrelation time induced by the finite bandwidth of the detection setup~\cite{supmat}. This limited sampling rate still enables an accurate estimation of the trajectory using a discrete time version of Eq.~\eqref{SME}~\cite{amini2011stability,rouchon2014efficient,supmat}. The quantum trajectories originating from the measurement records in Fig.~\ref{fig2}a,c are represented in Fig.~\ref{fig2}b,d. They present an erratic behavior coming from the randomness of the measurement backaction yet eventually converge towards $|g\rangle$ (south pole). This is similar to a random walk in the Bloch sphere with a step size that decreases to zero as the state approaches $|g\rangle$. Strikingly, the trajectories differ from the average one owing to a large enough measurement efficiency $\eta=24~\%$. If the detection was ideal ($\eta=1$) and without pure dephasing, the state would remain pure and the Bloch vector would evolve stochastically on the surface of the Bloch sphere. 
  
\begin{figure}
\includegraphics[scale=.21]{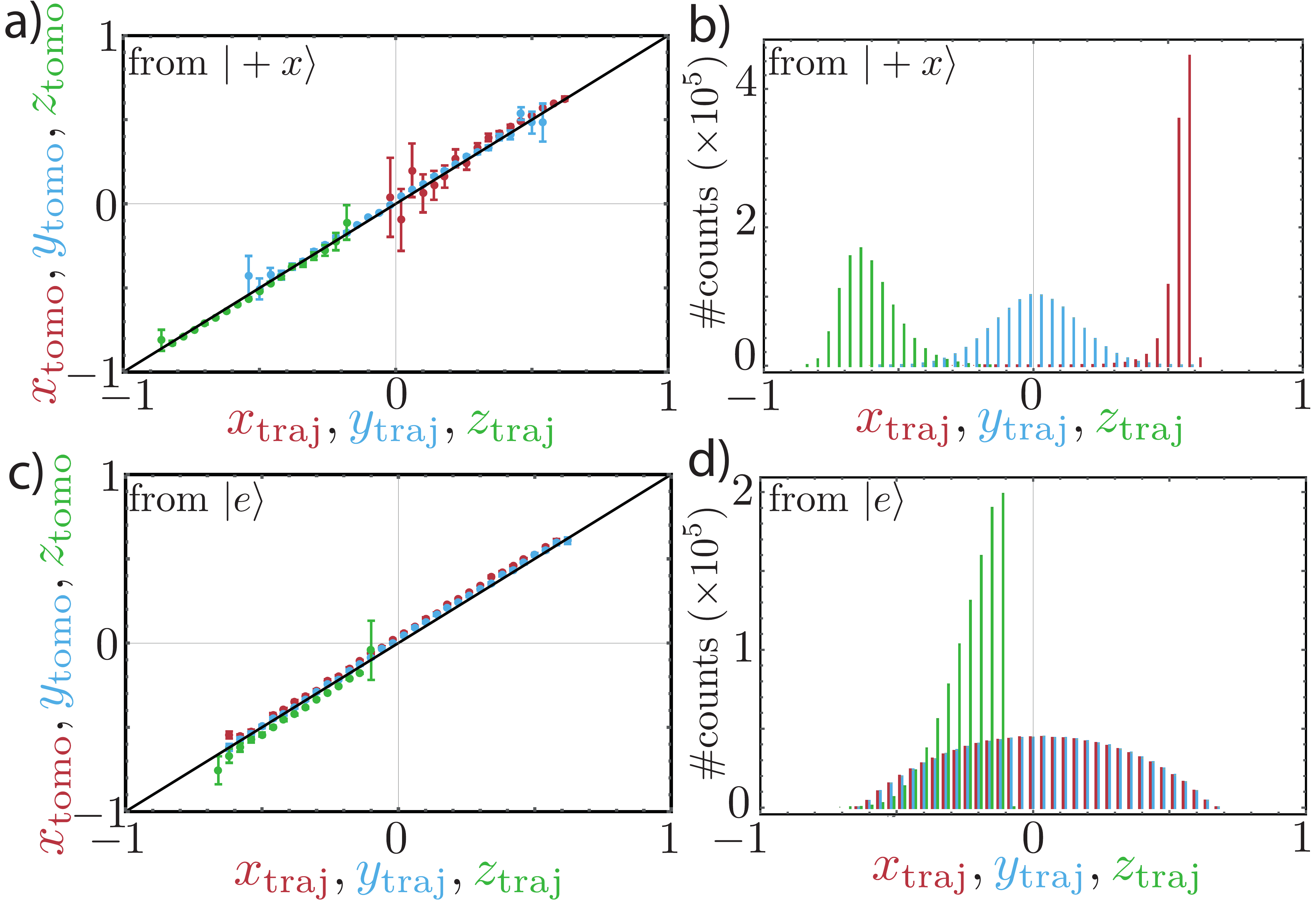}
\caption{\label{fig3} \textbf{Tomography versus quantum trajectories.} \textbf{a)} For each value of $x_\mathrm{traj}$, a red dot indicates the average value $x_\mathrm{tomo}$ of the final measurement of $\sigma_x$ on the subset of experiments starting from $|+x\rangle$ and for which the quantum trajectory ends up at time $T=4~\mu\mathrm{s}$ in a state such that $\langle\sigma_x\rangle_{\rho_T}= x_{\mathrm{traj}}~\pm~0.02$. Error bars represent the statistical uncertainty and the solid line has a slope 1.
Subsets with less than 40 experiments are not shown. The final projective measurement infidelity is corrected for~\cite{supmat}. Similar dots represent the case of $y$ (blue) and $z$ (green). \textbf{b)} Distribution of the final Bloch coordinate $i_\mathrm{traj}$ of the $10^6$ quantum trajectories starting in $|+x\rangle$ for $T=4~\mu\mathrm{s}$, and followed by a $\sigma_i$ projective measurement. The color code and the bin width match those in Fig.~\ref{fig3}a. \textbf{c-d)} Same as \textbf{a-b} for experiments starting in $|e\rangle$.}
\end{figure}

\section{Validation of the quantum trajectories by independent measurements}

\label{sectionvalidation}

Using Eq.~\eqref{SME}, we are able to reconstruct the quantum trajectories corresponding to an initial state at time $0$ and a fluorescence record between $0$ and $T$. By nature, the final state $\rho_T$ encodes the statistics of any measurement that would take place at time $T$. In order to test the pertinence of this prediction, one could use the very same fluorescence measurement at time $T$ and check whether or not Eq.~(\ref{dIdQ}) is satisfied on average. Here, we realize a more thorough test by performing independent measurements for $T$ ranging from 1 to $10~\mu\mathrm{s}$. A projective measurement of the qubit is thus performed by measuring its effect on a cavity mode~\cite{supmat}, following a $\pi/2$ pulse around  $\sigma_y$  or around $\sigma_x$, or no pulse at all (Fig.~\ref{fig1}b). The experiment is repeated one million times for each final measurement of $\sigma_x$, $\sigma_y$ or $\sigma_z$, for each preparation and for each final time $T$. 

 \begin{figure*}
\includegraphics[scale=.28]{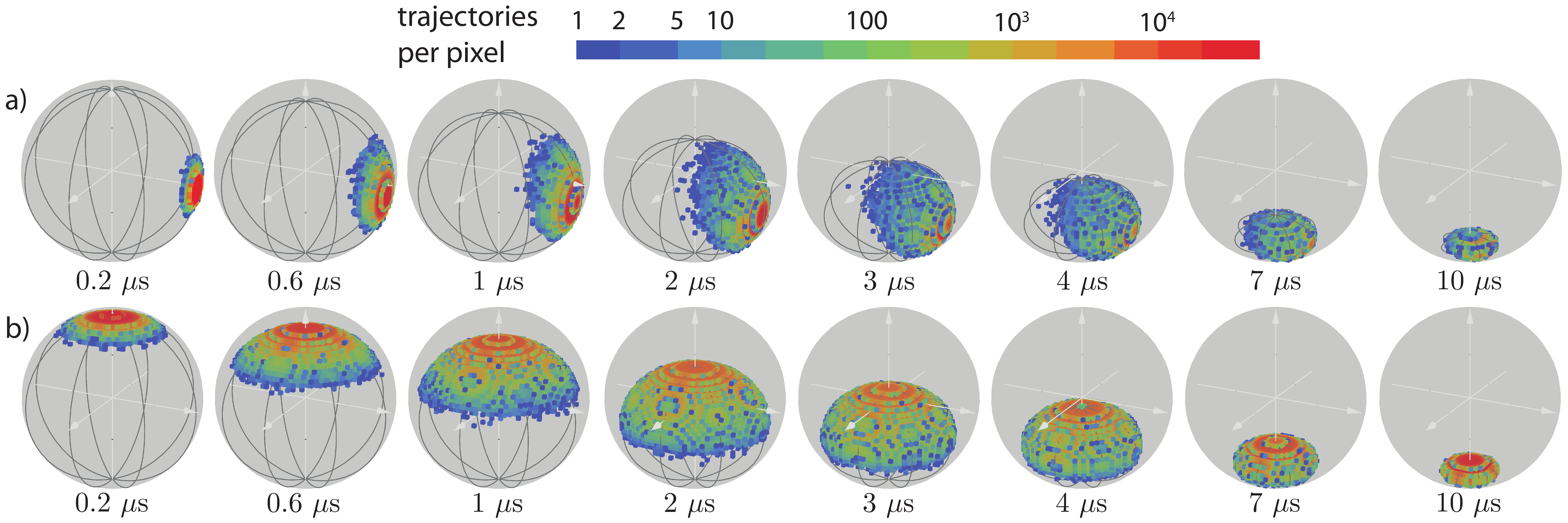}
\caption{\label{fig4} \textbf{Statistics of quantum trajectories.} Distributions of the qubit states along $10~\mu\mathrm{s}$-long trajectories for a qubit initially in $|+x\rangle$ (\textbf{a}) and $|e\rangle$ (\textbf{b}). The number of trajectories reaching each  cubic cell of side 0.04 is encoded in color, out of a total of 3 millions. White arrows go from $-1$ to $+1$ along $\sigma_{x,y,z}$ and the Bloch sphere is colored in gray. Meridians of the spheroid spanned by Eq.~\eqref{spheroid_eq} are also shown.}
\end{figure*}
In the case of a final measurement of $\sigma_x$, we then select the subset of all the realizations whose final predicted Bloch coordinate $\langle\sigma_x\rangle_{\rho_T}$ ends up within $2~\%$ of a given value $x_{\mathrm{traj}}$. The test then consists in comparing $x_{\mathrm{traj}}$ to the mean value $x_{\mathrm{tomo}}$ of the final $\sigma_x$ measurement outcome on that subset. The tomography results match the predictions of Eq.~\eqref{SME} as can be seen in  Fig.~\ref{fig3}a, where we plot in red $x_{\mathrm{tomo}}$ as a function of $x_{\mathrm{traj}}$ for the $T=4~\mu\mathrm{s}$-long trajectories starting in $|+x\rangle$.  Similar measurements are shown for the $\sigma_y$ and $\sigma_z$ measurements in the same figure. The error bars represent the statistical uncertainty on $x_{\mathrm{tomo}}$ corresponding to the limited number of selected realizations for a given $x_\mathrm{traj}$ plotted in Fig.~\ref{fig3}b~\cite{supmat}. The same verification is realized on experiments starting from $|e\rangle$ (Figs~\ref{fig3}c,d). Slight constant offsets between the predicted (\textit{traj}) and tomography (\textit{tomo}) values appear. The fact that these offsets are larger when the qubit starts in $|e\rangle$ than in $|+x\rangle$ indicates that they originate from systematic errors in the initial qubit preparation. Note that there is no fit parameter in the model beside $\eta$~\cite{six2015parameter}. The agreement between the states reconstructed using the fluorescence signal and the ones measured through tomography is also good for all other considered times $T$ (not shown here).

\section{Statistics of quantum trajectories}

Now that the approach is validated, the experiment can be used to probe the statistics of quantum trajectories. Fig.~\ref{fig3}b,d represent the distribution of predicted states at time $T=4~\mu\mathrm{s}$. The average of each Bloch coordinate matches its expected value after a relaxation of $4~\mu\mathrm{s}$.
The spread of the distributions comes from measurement backaction. Starting from $|e\rangle$ (Fig.~\ref{fig3}d), the qubit state has no defined initial phase, and this symmetry is preserved: $\langle\sigma_x\rangle_{\rho_T}$ and $\langle\sigma_y\rangle_{\rho_T}$ remain $0$ on average. However, at the single realization level, coherences develop in time by spontaneous symmetry breaking due to the inherent randomness of the measurement process. This is in sharp contrast with the quantum trajectories obtained by continuous dispersive measurement of $\sigma_Z$~\cite{murch2013observing} when starting from an energy eigenstate. 

The statistics of quantum trajectories can be better understood in the Bloch sphere representation. In Fig.~\ref{fig4}, is represented the distribution of the qubit states at various times $t$ for $10~\mu\mathrm{s}$-long trajectories, for a qubit starting in $|+x\rangle$ (a) or $|e\rangle$ (b).  Starting from a single point, the state distribution progressively spreads out and collapses down to $|g\rangle$ at long times. Note that at the first times in the evolution, the distribution spread in the Bloch sphere is larger when starting from $|e\rangle$ than from $|+x\rangle$. This illustrates that the measurement backaction associated with spontaneous emission is as strong as the qubit excitation is large. Strikingly, at each time in the evolution, all quantum states seem to belong to the same shell in the Bloch sphere, independently of the initial state.

An analytical expression of this surface can in fact be derived when neglecting dephasing ($\gamma_\phi=0$ in Eq.~\eqref{SME}). Let us introduce the variable 
\begin{equation}
\alpha=1+\frac{1}{2}\frac{S_L}{p_e^2}\geq1,
\end{equation}
where, $S_L=1-\mathrm{Tr}(\rho_t^2)$ is the linear entropy and $p_e=(1+\langle\sigma_z\rangle_{\rho_t})/2$ is the probability to find the qubit excited. It can then be shown~\cite{supmat} that $\alpha(t)$ 
evolves deterministically, independently of the heterodyne fluorescence record, following
\begin{equation}
\alpha(t)=\eta+(\alpha(0)-\eta)e^{\gamma_1t}.
\end{equation}
Remarkably, the qubit state does not diffuse stochastically in the volume of the Bloch sphere. Rather, it is restricted to the surface determined by the value $\alpha(t)$ whose characteristic equation reads
\begin{equation}
\alpha(x^2+y^2)+\alpha^2 (z+1-\frac{1}{\alpha})^2=1.\label{spheroid_eq}
\end{equation}
The surface is thus a spheroid going through the south pole of the Bloch sphere $|g\rangle$. Starting from a pure state at $\alpha=1$, the spheroid shrinks from the Bloch sphere itself towards the south pole as $\alpha$ rises in time. As shown in Fig.~\ref{fig4}, it is in good agreement with the measured distributions. The small thickness~\cite{supmat} of the shell in Fig.~\ref{fig4} originates from pure dephasing alone. Note that in the ideal case $\eta=1$, the trajectories would evolve on the sphere as $\alpha(t)=1$ is then constant.

On the spheroid, the evolution is still stochastic and depends on the particular realization of the heterodyne measurement of fluorescence. Yet, it is possible to identify integral quantities of the measured quadratures $\{\mathrm{d}I_t,\mathrm{d}Q_t\}$ alone that relate directly to the position on the spheroid, hence avoiding to solve numerically Eq.~\eqref{SME}. For that purpose, the position of a state on the spheroid is parametrized by the variables $\xi_x(t)=\frac{\langle\sigma_x\rangle_{\rho_t}}{\langle\sigma_z\rangle_{\rho_t}+1}$ and $\xi_y(t)=\frac{\langle\sigma_y\rangle_{\rho_t}}{\langle\sigma_z\rangle_{\rho_t}+1}$. We show that~\cite{supmat}
\begin{equation}
\left\{\begin{array}{c}
\xi_x(t)=\xi_x(0)e^{\gamma_1t/2}+\sqrt{\frac{\eta\gamma_1}{2}}\int_0^te^{\gamma_1(t-\tau)/2}\mathrm{d}I_\tau\\
\xi_y(t)=\xi_y(0)e^{\gamma_1t/2}+\sqrt{\frac{\eta\gamma_1}{2}}\int_0^te^{\gamma_1(t-\tau)/2}\mathrm{d}Q_\tau
\end{array}\right. .
\end{equation}
These expressions can be tested using the same tomography measurements as in Fig.~\ref{fig3} and lead to a similar agreement~\cite{supmat}.

For some trajectories in Fig.~\ref{fig4}a starting from $|+x\rangle$, the probability for the qubit to be excited increases in time as $\langle\sigma_z\rangle_{\rho_T}$ temporarily takes positive values. This is counterintuitive as the average energy increases for a subset of experiments that can be postselected based on the field emitted during decay~\cite{Bolund2014}. Yet there is no paradox here since the initial state is not an energy eigenstate. On a single experiment, the information extracted by the fluorescence monitoring may indeed resolve this quantum uncertainty and favor the excited state. This observation makes explicit the inadequacy of a reasoning in terms of emitted photons in the case of heterodyne or homodyne detection. In contrast, any signal detected by a photo counter would indeed be associated with decreasing qubit energy. A trajectory with increasing $\langle\sigma_z\rangle$ is shown in Fig.~\ref{fig2}a and results from a measurement record that starts with negative values of $\mathrm{d}I_t$ over some time interval while the qubit state is still close to $|+x\rangle$.

\section{Conclusion}

Quantum state diffusion was first introduced as a description of what an open quantum system undergoes from the viewpoint of its environment~\cite{Gisin1992}. This experiment, which implements the proposal of Wiseman and Milburn in 1992~\cite{Wiseman1993,wiseman2009quantum}, illustrates how it can in fact be understood as the dynamics of a system conditioned to a continuous measurement record. In that respect, it helps to better understand relaxation and decoherence. The use of a Josephson Parametric Converter was instrumental in order to reach a high enough measurement efficiency so that measurement backaction is visible and quantum trajectories depart from their average. Importantly, it was possible to validate the reconstructed quantum trajectories by an independent quantum state tomography. The experiment thus demonstrates how the sole continuous monitoring of a relaxation channel can lead to various quantum states and even produce superpositions of classical ones. It opens the way to experiments in which the field emitted during relaxation is used as an input of a feedback controller able to stabilize any state of a qubit~\cite{Hofmann1998}. This would complement the toolbox of quantum error correction, which is key to the development of quantum computing. Finally, the possibility to observe statistics of quantum trajectories should lead to interesting perspectives in the field of thermodynamics of quantum information.

\begin{acknowledgments} We thank Alexia Auff\`eves, Maxime Clusel, Pascal Degiovanni, Michel Devoret, Emmanuel Flurin, Fran\c cois Mallet and Vladimir Manucharyan for fruitful discussions. Nanofabrication has been made within the consortium Salle Blanche Paris Centre. This work was supported by the EMERGENCES grant QUMOTEL of Ville de Paris and by the IDEX program ANR-10-IDEX-0001-02 PSL $^*$.\end{acknowledgments}

\FloatBarrier

\bibliographystyle{landry}

\end{document}